\documentstyle[epsfig]{aipproc}
\begin{document}

\pagestyle{plain}

\title{Two-Photon Physics at RHIC: Separating Signals from Backgrounds}

\author{Joakim Nystrand and Spencer Klein}
\address{Lawrence Berkeley National Laboratory\\
Berkeley, California 94720, U.S.A.}

\maketitle

\vspace{-6.0cm}
\begin{flushright}
LBNL--41111 \\
November, 1997
\end{flushright}

\vspace{6.2cm}

\centerline{\small\it Presented at Hadron '97, Brookhaven National Laboratory, August 25--30, 1997}

\vspace{-3.5cm}

\begin{abstract}
This presentation will show the feasibility of studying 
two--photon interactions in the STAR experiment at RHIC. 
Signals, detection efficiencies, backgrounds, triggering and
analysis techniques will be discussed.
\end{abstract}

\vspace{0.3cm}
\thispagestyle{plain}

In the Relativistic Heavy--Ion Collider\cite{RHIC} (RHIC), under construction 
at the  Brookhaven National Laboratory and scheduled to begin operation in 
1999, beams of heavy--ions as massive as gold (A=197) will 
collide at center--of--mass energies of 100+100~A~GeV. The main purpose of 
RHIC is to study central nucleus--nucleus 
collisions in order to investigate the properties of nuclear matter 
under extreme conditions. Of particular interest will be to search for a 
possible transition from ordinary hadronic matter to a quark--gluon plasma. 
However, due to the strong electromagnetic and nuclear fields created when 
two ions sweep past each other, many interesting interactions occur when the 
nuclei miss each other.
This includes two--photon
interactions, $\gamma$--Pomeron and perhaps Pomeron--Pomeron interactions.
Studying the production of hadrons in $\gamma \gamma$ interactions can
probe the quark content of exotic hadronic states.
Measurements of $\gamma$--Pomeron or Pomeron--Pomeron interactions will reveal
how the Pomeron couples to the nucleus.
The Peripheral Collisions Program within the STAR (Solenoidal Tracker At 
RHIC) collaboration\cite{STAR} will study 
interactions of this type. This talk will discuss the feasibility of 
extracting a signal from two--photon interactions and separating it from 
background.

\vspace{-0.3cm}
\section*{Two--Photon Interactions}

\vspace{-0.3cm}
Two--photon interactions have been previously studied at  $e^+e^-$--colliders.
Heavy--ion colliders are particularly attractive for investigating these 
reactions because the electromagnetic fields of the protons add 
coherently. 

\begin{figure}[t!] % fig 1

\begin{center}
\begin{minipage}{2.4in}
\epsfig{file=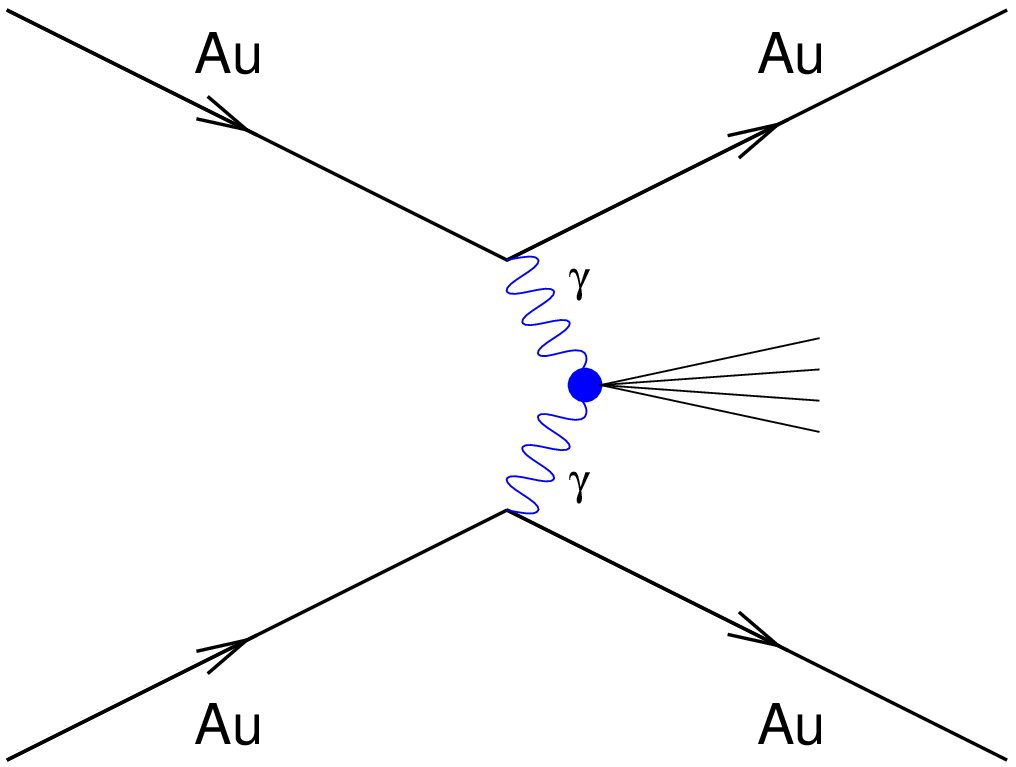,height=1.8in,width=2.4in}
\end{minipage}
\hspace*{0.4in}
\begin{minipage}{2.4in}
\epsfig{file=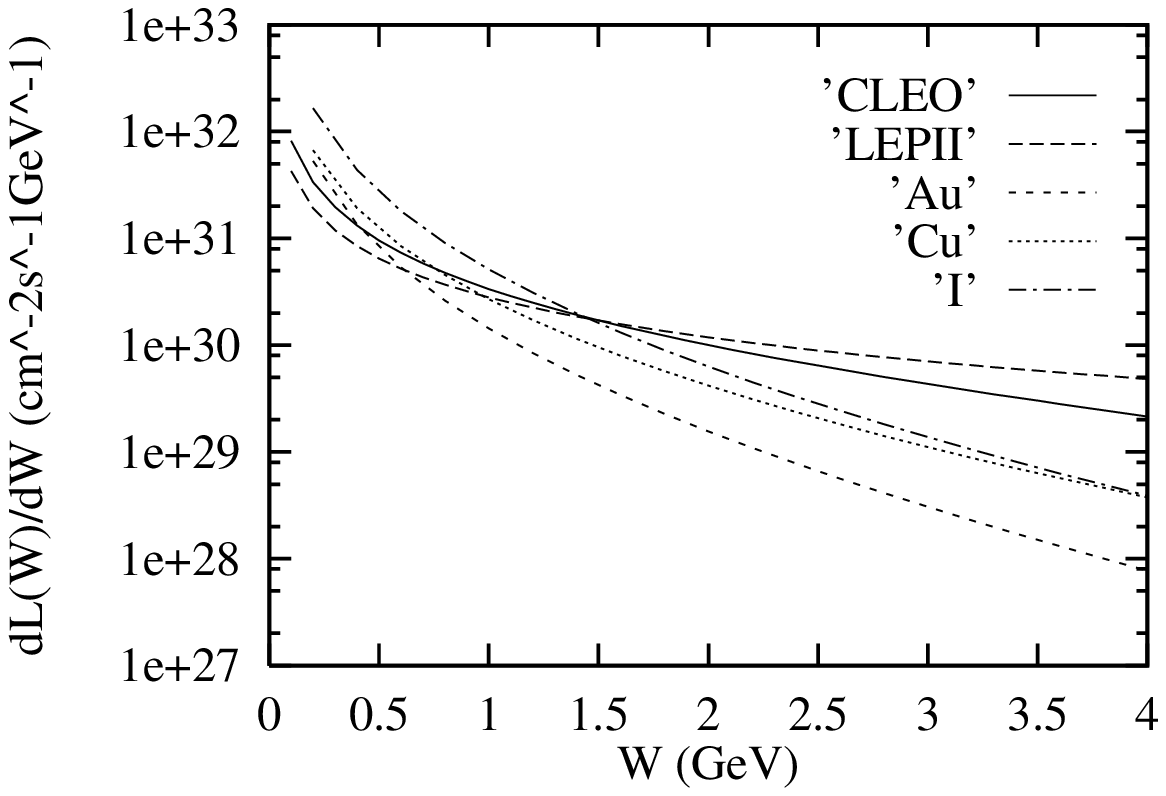,height=1.8in,width=2.4in}
\end{minipage}
\end{center}

\vspace{10pt}
\caption{Left: A schematic illustration of a two--photon interaction 
$Au+Au \rightarrow$ \protect \linebreak $Au+Au+X$. Right: The two--photon 
luminosity at RHIC compared with that at LEP and CLEO.}
\label{fig1}

\end{figure}

In the Weizs\"{a}cker--Williams method the electromagnetic field of a 
relativistic, charged particle is treated as a pulse of photons. For a
spherically symmetric charge distribution with radius $R$, the photon energy 
spectrum integrated over impact parameter from $R$ to infinity is:
\begin{equation}
   E_{\gamma}\frac{dN}{dE_{\gamma}} = \frac{2Z^{2}\alpha}{\pi} 
   \left\{ x K_{0}(x) K_{1}(x) - \frac{1}{2} x^{2} \left( K_{1}^{2}(x) - 
   K_{0}^{2}(x) \right) \right\}
   \label{dndE}
\end{equation}
where $x = E_{\gamma} R / \gamma$ and $E_{\gamma}$ is the photon energy 
\cite{Jackson}.

When calculating the $\gamma \gamma$--luminosity in heavy--ion collisions, 
it is necessary to exclude the region of nuclear overlap ($b<R_1+R_2$), as 
in this region hadronic interactions dominate. The equivalent two--photon 
luminosity is\cite{Baur,Cahn,Hencken}:
\begin{equation}
   {\cal L}_{\gamma \gamma}(E_{\gamma}^1,E_{\gamma}^2) \propto
   f(E_{\gamma}^1) f(E_{\gamma}^2) - 
   \Delta F(E_{\gamma}^1,E_{\gamma}^2) 
\end{equation}
where $f(E_{\gamma})$ is the single photon distribution (Eq.~1) and 
$\Delta F(E_{\gamma}^1,E_{\gamma}^2)$ is a function which takes into account 
the overlap effect:
\begin{equation}
   \Delta F(E_{\gamma}^1,E_{\gamma}^2) = 4\pi \int_{R}^{\infty} b_1 db_1
   \int_{b_{min}}^{b_1+2R} b_2 db_2 \frac{d^2 f}{db_1^2}
   \frac{d^2 f}{db_2^2} 
   \arccos(\frac{b_1^2+b_2^2-4R^2}{2b_1b_2})
\end{equation}

The two--photon luminosity thus essentially scales as $Z^4$, and this is a 
major advantage of heavy--ion colliders. Figure~1 shows the luminosity for 
various nuclear systems at RHIC\cite{Spencer,StarNote}.
The two--photon luminosity will be competitive with 
that at $e^+e^-$--colliders, such as CLEO and LEPII, up to a center--of--mass 
energy of about 1.5~GeV.

\vspace{-0.2cm}
\section*{The STAR Experiment at RHIC}

\vspace{-0.3cm}
The main detector of the STAR experiment at RHIC will be a large 
($\sim$50~m$^3$), cylindrical Time Projection Chamber (TPC). The 
TPC will track and reconstruct momenta of charged particles 
in the pseudo--rapidity interval $-2 < \eta < 2$. Two smaller 
forward time projection chambers (FTPC) will cover the interval 
$2.5 < \mid \eta \mid < 3.75$. 

One of the major challenges in studying two--photon interactions in a 
nuclear experiment will be to trigger on these interactions without 
collecting too many background events.
The STAR trigger system is well suited for this purpose. The main TPC 
will be covered with an array of 240 plastic scintillators, which 
will measure the charged particle multiplicity in the interval 
$\mid \eta \mid < 1.0$ and which will serve as trigger detectors. 
The TPC anode wires will provide multiplicity information for 
$1.0 < \mid \eta \mid < 2.0$ for the trigger.

The raw data from 
the scintillator array and the wire chambers will be available in less 
than 1~$\mu$s. More accurate multiplicity information will be available after
about 100~$\mu$s. In the highest trigger level, after 10~ms, some tracking and 
momentum information from the TPC will also be available. This will enable 
the primary vertex to be located and the $\sum p_{T}$ to be 
determined. The different trigger levels and their respective time--scales
are summarized in Table~1.

Central Au+Au events will be recorded at a rate of 
about 1~Hz. The data acquisition system will, however, be flexible 
enough to allow the recording of smaller events, such as two--photon 
interactions, at a higher rate.

\vspace{-0.3cm}
\begin{center}
\begin {table} [thb] \begin{center} 
\begin{tabular} {crc} \hline
Trigger Level & Decision Time     & Selection                                \\ \hline
   0          & $\sim$   1 $\mu$s &  $2 \leq n_{ch} \leq 5$, event topology        \\
   1          & $\sim$ 100 $\mu$s & $n_{ch}=$2 or 4, event topology                \\
   3          & $\sim$  10 ms     & $\sum p_{T} \leq$ 100 MeV/c, vertex position\\ \hline
\end{tabular} \end{center}
\label{trigger1}
\caption{A summary of the trigger levels in STAR and cuts applied in the
peripheral collisions analysis. Level 2 is not used for peripheral collisions.}
\end{table}
\end{center}

\vspace{-1.2cm}
\section*{Signal and Background Rates}

\vspace{-0.3cm}
The production and background rates of the following three systems have been 
studied: a lepton pair ($\mu^+\mu^-$), a single meson 
decaying into 2 charged particle ($f_2(1270) \rightarrow \pi^+ \pi^-$)  
and a meson pair decaying into 4 charged particles 
($\rho^0\rho^0  \rightarrow \pi^+ \pi^- \pi^+ \pi^-$). These systems should 
thus be representative of a wide selection of two--photon final states. All three 
systems were simulated using STARLight, a two--photon Monte--Carlo 
generator\cite{Spencer,StarNote}. In STARLight, photon transverse momenta
are generated with a Gaussian form factor with width $\hbar c/R$.

Four sources of background have been considered:
peripheral (hadronic) nucleus--nucleus collisions, beam--gas interactions, 
$\gamma$--nucleus interactions, and cosmic rays.  
Cosmic rays and  beam--gas reactions are a problem mainly 
at the trigger stage. Off--line, it will be possible to 
accurately find
the primary vertex and thence minimize the cosmic ray 
contribution and reduce the background beam--gas interactions.

Background interactions have been simulated with  several different 
models. For hadronic nucleus--nucleus collisions and beam--gas interactions, 
two standard nucleus--nucleus Monte--Carlo models, FRITIOF 7.02\cite{Fritiof} 
and VENUS 4.12\cite{Venus}, have been used. Photonuclear interactions have 
been simulated with the model DTUNUC 2.0\cite{dtunuc}. For cosmic 
rays, the Monte--Carlo HemiCosm has been used\cite{BaBar}.

\begin{center}
\begin {table} [t!] \begin{center}
\begin{tabular} {ccccccc} \hline
\footnotesize
Trigger         & \multicolumn{2}{c}{Hadronic A+A} & \multicolumn{2}{c}{Beam--Gas} &
$\gamma$+A       & Cosmic Rays \\
Level   & FRITIOF & VENUS & FRITIOF & VENUS & DTUNUC & HemiCosm \\ \hline
0       & 19      & 21    & 53      & 53    &  63    &  30      \\
1       &  7      &  9    & 27      & 28    &  37    &  30      \\
3       &  0.1    &  0.2  &  0.2    &  0.3  &   1.9  &   0.6    \\ \hline
\end{tabular}
\label{trigger2}
\caption{Trigger Rates (in Hz) of the various background processes discussed in the 
text.}
\end{center}
\end{table}
\end{center}
\normalsize

\vspace{-0.7cm}
The background trigger rates obtained using these models are shown in Table~2.
The trigger selection used at each level is summarized in Table~1.
At level 3, reactions are required to occur inside the interaction 
diamond, $\mid z \mid \leq 20$~cm, and have 
$\mid \sum \overline{p}_{T} \mid \leq 100$~MeV/c. The largest background component is 
photonuclear interactions. The trigger rates can be compared to the trigger rate from 
muon--pairs from $\gamma \gamma$ interactions, 4.7~Hz. The ratio of signal to 
background for $\mu$--production will thus be roughly 2:1, already at the trigger 
level.
 
In order to separate signals from background, cuts have been 
developed which utilize the characteristics of two--photon interactions. 
The most important of these cuts are: (a) Multiplicity: Many two--photon 
reactions lead to a final state with either 2 or 4 charged particles, and 
in the analysis it is required that no additional particles are present in 
the event. (b) Transverse momentum: The summed transverse momentum 
($\mid \sum \overline{p}_{T} \mid$)  of the final state will be small 
($\sim \sqrt{2} \hbar c / R$). In the analysis presented here a cut of 
$\mid \sum \overline{p}_{T} \mid \leq$~40~MeV/c was used. 
(c) Center--of--mass rapidity: The rapidity distribution 
of the $\gamma \gamma$ system is centered around 0 with a fairly narrow width 
($y_{cm} \lesssim 1-2$). In the analysis a cut of $y_{cm} \leq$~1.0 was 
applied. 
A cut on $y_{cm}$ reduces the number of beam--gas and photonuclear 
interactions, since these are characterized by asymmetric particle emission 
relative to the Au+Au center--of--mass system. 

The expected integrated rates for $f_2(1270)$ and $\rho^0\rho^0$ near 
threshold (1.5 $\leq m_{\rho \rho} \leq$ 1.6~GeV/c$^2$) for one 
year of running is presented in Table~3. One year corresponds to $10^7$~s of 
beam--time at the RHIC design luminosity for Au+Au 
(${\cal L}= 2.0 \cdot 10^{26}$~cm$^{-2}$s$^{-1}$). For the rates in 
Table~3, a cut was also applied in invariant mass of the final state. For 
$f_2(1270)$, $m_{\pi \pi}$ was required to be within $\mid m_{\pi \pi} - m_{f_2} \mid \leq \Gamma$,
where $\Gamma$ is the natural width, and for $\rho^0\rho^0$ it was required
that $1.5 \leq m_{inv} \leq 1.6$~GeV/c$^2$. 

Both of these signals can be clearly separated from background.
The backgrounds estimated by FRITIOF and VENUS differ considerably for
hadronic A+A interactions. The reason for this is not
entirely understood. The two models are in general agreement regarding
multiplicity. The differences occur when cuts
are applied in for instance $\mid \sum \overline{p}_{T} \mid$. Of course,
neither of these models have been compared with data from heavy--ion
reactions at RHIC energies. The difference between the 
models can be taken as a measure of the systematic error in the estimate. 
As at the trigger level, the largest source of background seems to be 
photonuclear interactions.

\begin{center}
\begin {table} [t!] 
\begin{tabular} {lrr} \hline
                              & \multicolumn{2}{c}{System}           \\ 
                              & $f_2(1270)$      & $\rho^0\rho^0$ \\ \hline
Produced at RHIC              & 1,900,000        & 42,000         \\
Within STAR Acceptance        &   660,000        & 12,000         \\ \hline
Peripheral AA (FRITIOF/VENUS) &     1,000/5,000  &    200/800     \\
Beamgas (FRITIOF/VENUS)       &     1,000/1,000  &    100/100     \\
$\gamma$+A (DTUNUC)           &    24,000        &  1,200         \\ \hline
Total Background              &    26,000/32,000 &  1,500/2,100   \\ \hline
\end{tabular}
\label{f2}
\caption{Estimated signal and background rates (Events/year) for production of
$f_2(1270)$ and $\rho^0\rho^0$.}
\end{table}
\end{center}

\vspace{-1.5cm}
\section*{Conclusions}

\vspace{-0.3cm}
To summarize, we have shown that ultra--relativistic heavy--ion colliders 
provide high rates of $\gamma \gamma$ interactions, and that STAR
is well suited to study this type of physics. 
Using appropriate cuts, signals can be separated from background both at the
trigger and analysis level. 

\vspace{0.3cm}
\noindent We thank S.~Roesler for providing the DTUNUC Monte Carlo code and for 
valuable discussion. We also thank our colleagues in STAR for their support. This 
work was supported by the U.S.~DOE under contract DE--AC--03--76SF00098.

\end{document}